\documentclass[a4paper,11pt]{article}
\pdfoutput=1 

\usepackage{jinstpub} 

\title{ Using $^{22}$Na and $^{83{\rm m}}$Kr to calibrate and study the properties of scintillation in xenon-doped liquid argon}

\author[a,b]{Y.Y. Gan,}
\author[a]{M.Y Guan,}
\author[a,1]{Y.P. Zhang,\note{Corresponding author.}}
\author[a,b]{P. Zhang,}
\author[a,b]{C.G. Yang,}
\author[a,b]{Q. Zhao,}
\author[a,b]{Y.T. Wei}
\author[a,b]{W.X. Xiong}

\affiliation[a]{Institute of High Energy Physics, Chinese Academy of Sciences, Yuquan Road, Beijing,100049, China}
\affiliation[b]{University of Chinese Academy of Sciences, Yuquan Road, Beijing, 100049, China}

\emailAdd{ypzhang1991@ihep.ac.cn}

\abstract{We have measured the properties of scintillation light in liquid argon doped with xenon concentrations from 165 ppm to 10,010 ppm using a $^{22}$Na source. The energy transfer processes in the xenon-doped liquid argon are discussed in detail, and a new waveform model is established and used to fit the average waveform. The time profile of the scintillation photon in the xenon-doped liquid argon and of the TPB emission are presented. The quantities of xenon-doped are controlled by a Mass Flow Controller which is calibrated via a Redusial Gas Analyzer to ensure that the xenon concentration is accurate. In addition, a successful test of $^{83{\rm m}}$Kr as a calibration source has been implemented in the xenon-doped liquid argon detector for the first time. By comparing the light yield of the $^{22}$Na and $^{83{\rm m}}$Kr, it can be concluded that the scintillation efficiency is almost same over the range of 41.5 keV to 511 keV.}

\keywords{Noble-liquid detectors, Liquid argon, Scintillators, Redusial Gas Analyzer}

\begin{document}
\maketitle
\flushbottom

\section{Introduction}

Liquid argon is widely used as the detection medium in experiments hunting rare events, in particular, for searching dark matter~\cite{WArP,DEAP-3600,DarkSide}, detecting neutrino~\cite{DUNE, ArgoNeuT} and the measurement of coherent elastic neutrino-nucleus scattering (CN${\nu}$NS)~\cite{COHERENT}. The advantages of liquid argon lie in the following aspects:

1)The high scintillation efficiency and self-transparency to the scintillation light~\cite{Yield};

2)The timing components of liquid argon scintillation light provide powerful signal pulse shape discrimination (PSD) capabilities to separate nuclear recoils (NR) and electronic recoils (ER)~\cite{PSD};

3)For a dual-phase detector composed of the noble-gas liquid, the signal induced by the primary ionization would be detected through the effect of proportional electroluminescence in the gas phase. This signal becomes of paramount importance to low-threshold rare event experiments, such as low-mass (< 10 GeV) dark matter search experiments~\cite{sub-GeV} and particular CN${\nu}$NS experiments~\cite{S2};

The disadvantage of liquid argon is that the scintillation light belongs to vacuum ultraviolet light, which has a center wavelength of 128 nm with a FWHM of about 10 nm~\cite{waveformlength} and is difficult to be detected by commercial photodetectors. The usual way to solve this problem is to use wavelength shifters (WLS) to convert the wavelength into visible range, for example coating a thin film of TetraPhenyl Butadiene (TPB)~\cite{TPB,TPB-2} on the detector walls, on PMT windows, or on optically transparent plates in front of PMTs.

Many previous literatures which study on doping liquid argon with small amounts of xenon have shown that xenon dopant can improve the light yield and shorten the overall duration of the waveform~\cite{1993-1,1993-2,1996,2014,2019}. It is also known that adding xenon to argon leads to a strong modification of the emission spectrum. The peak wavelength of the emission spectrum shifts from 128 nm of argon scintillation to 176 nm of xenon scintillation which can be detected by commercial photodetectors. Therefore, xenon-doped in liquid argon could work as a volume-distributed WLS~\cite{1993-3,2017} which is expected to provide better positional reconstruction capability since the re-emission occurs in the point of interaction.

For different experimental groups, the proportions of increased light yield varying from a slight increase to 2 times are inconsistent, which may be caused by the difference in detector performance. A shorter waveform duration mainly caused by energy transfer from the long-lived argon triplet excimers to xenon which has a much shorter decay time, is also desirable for much shorter detector dead time. This phenomenon offers xenon-doped liquid argon a potential application in high count rate detectors, such as Positron Emission Tomography (PET)~\cite{PET}. The experimental studies ~\cite{2019,1993-4,2017-1} also indicate the evidence of fast component re-emission in liquid argon doped at high xenon concentrations. 


In many previous xenon doping experiments, the xenon concentration was determined mainly in two ways, one of which is to use a evacuated small chamber with the known volume, and then fill it with a certain pressure of xenon. Therefore, the quantities of xenon dopant are determined by the pressure variation of the pressure gauge at room temperature. The relative error of the mixture prepared with the above described procedure is relatively large and up to about 50$\%$~\cite{2019,2017}. The other way is using gas chromatography or mass-spectrometry to accurately measure xenon concentration in argon xenon mixtures, which has a relatively high accuracy. However, the relative error also depends on the experimental processes and operation methods, and it can be up to about 10$\%$~\cite{2019}. 

In our work, a Mass Flow Controller (MFC) which has higher accuracy than the above method using the pressure gauge is applied to determine the quantities of xenon-doped. The MFC is calibrated for specific gas, especially xenon and argon. Since xenon freezes at the temperature of liquid argon, xenon gas needs to be diluted and mixed with enough argon in the gas phase before doping xenon in liquid argon. Therefore xenon gas has to slowly flow into the argon circulation line through the MFC, and then is liquefied together with argon to prevent freezing inside the tubing and cold head. In order to ensure that the xenon concentration measured by the MFC is accurate after sequential xenon doping, the independent measurements of relative argon xenon gas ratio at different concentration calculated by the MFC were done with a Redusial Gas Analyzer (RGA)~\cite{RGA}. In our future works, the RGA will be used to directly measure the ratio of xenon to argon after evaporation to ensure that the xenon will not freeze on the cavity wall.

In order to study the waveform shape dependence at different xenon concentrations, 511 keV $\gamma$-rays emitted from the $^{22}$Na radioactive source is used to calibrate the detector. $^{83{\rm m}}$Kr is considered to be an ideal calibration source for liquid noble gas detector because it is a low-energy source and can be easily injected into detectors without any contamination~\cite{2015,2019-1,Kr83}. The $^{83{\rm m}}$Kr source comes from decay of $^{83}$Rb, which can be obtained by high speed proton hitting natural krypton gas. $^{83}$Rb decays to $^{83{\rm m}}$Kr with a half-time of 86.3 days. Then $^{83{\rm m}}$Kr decays to the ground state with a half-time of 1.83 hours, emitting 32.1 keV and 9.4 keV conversion electrons. After doping 10010 ppm xenon, we calibrate the detector with $^{83{\rm m}}$Kr source which is applied in xenon-doped liquid argon detector for the first time.
 
\section{Experimental setup}
\subsection{Xenon doping and calibration system}

  \begin{figure}[htbp]
  \centering
  \includegraphics[width=9cm]{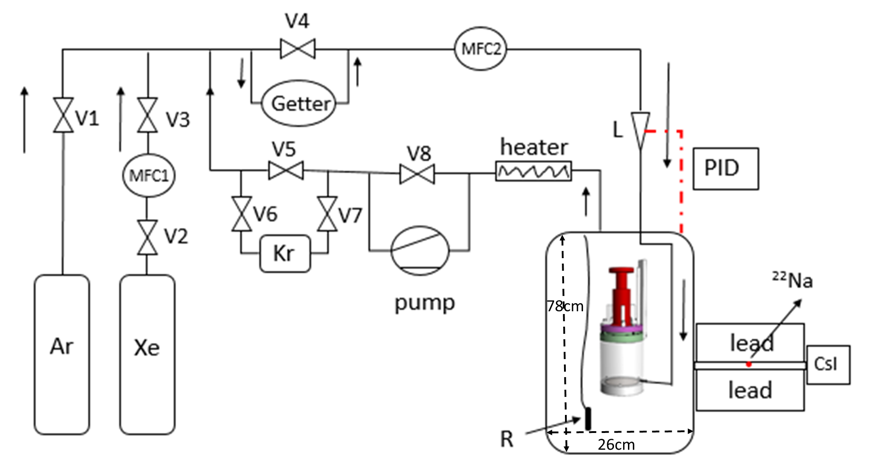}
  \caption{\label{fig:measurement_system} Schematic view of the xenon doping and calibration system. V1$\sim$V5 are the vacuum diaphragm valves, L is a liquid nitrogen cryocooler, and R is a heating resistor for vaporizing liquid argon. Use the MFC1 and the MFC2 to measure the amount of xenon and argon filled into dewar vessel, respectively. 
  }
  \end{figure} 
  
\begin{figure}[htbp]
  \centering
  \includegraphics[width=7cm]{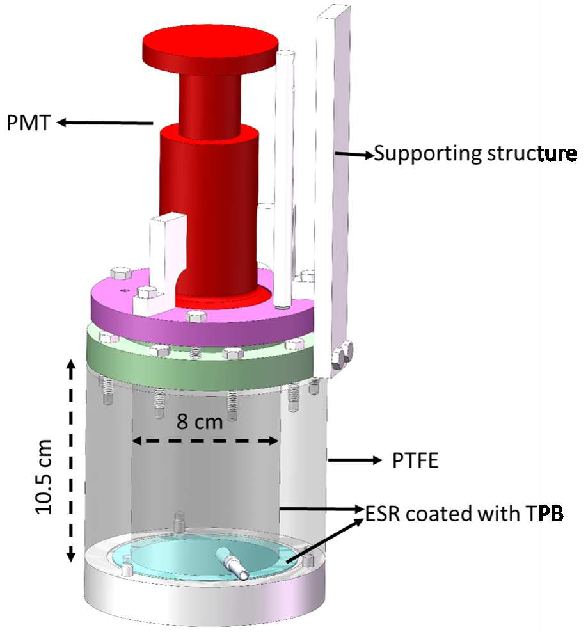}
  \caption{\label{fig:TPC} Three-dimensional diagram of the single phase detector.
  }
  \end{figure}

The xenon doping and calibration system shown in Fig.~\ref{fig:measurement_system} consists of three parts, namely gas-handling, detector and calibration source. With this system, the operation processes of argon gas filling, liquification and circulation purification, xenon doping into liquid argon, stable operation of the detector and calibration of the detector, could be achieved. The entire system is evacuated to measure the leak rate about 1 $\times$ 10$^{-9}$ Pa$\cdot$m$^{3}$/s using helium mass spectrometer leak detector before measurement. By controlling valves, commercial high purity argon(99.999$\%$ purity) gas and xenon (99.9999$\%$ purity) gas are further purified by getter (Simpure 9NG) to remove N$_{2}$, O$_{2}$ and other impurity gases, and then are liquefied by a liquid nitrogen cryocooler to flow into the dewar vessel. The single phase detector, shown in Fig.~\ref{fig:TPC}, composed of a polytetrafluoroethylene (PTFE) sleeve with a dimension of 8 cm in diameter and 10.5 cm in height, a 3 inch Photomultiplier tube (PMT, Hamamatsu-R11065) and some electronic components are immersed in xenon-doped liquid argon. TPB with a thickness of about 183 $\mu$g/cm$^{2}$ is coated on the window of the PMT, and the PMT is placed at the top of the sleeve to detect the scintillation. In order to improve the detection efficiency of scintillation, A layer of enhanced specular reflector film (ESR) is placed on the surface of the PTFE sleeve to enhance the reflectivity. The inner surfaces of ESR are coated with TPB to shift the 128 nm or 176 nm scintillation light to 420 nm which can be detected by the PMT. The thickness of the TPB coated on the ESR placed on the side wall and bottom of the sleeve is 283 $\mu$g/cm$^{2}$ and 210 $\mu$g/cm$^{2}$, respectively. 

Before xenon doping, a total of 13.3 kg of argon is filled into dewar vessel at 20 L/min flow rate. In addition, the MFC that controls the xenon inflow rate is set to about 10 ml/min, which is much smaller than that of argon to prevent xenon from freezing. Moreover, the slow control software designed based on labview programming language can monitor the temperature of liquid, pressure of the system and xenon doping inflow rate in real time. The temperature and pressure during the time of data acquisition were controlled within the range (90.5 $\pm$ 0.3) K and (47 $\pm$ 3) kPa respectively. The system stability was closely monitored throughout the series of tests. Five different concentrations of xenon argon mixtures were prepared, with concentrations of (165 $\pm$ 16) ppm, (500 $\pm$ 28) ppm, (1002 $\pm$ 46) ppm, (5005 $\pm$ 190) ppm and (10010 $\pm$ 370) ppm. The errors are calculated in consideration of the MFC accuracy, electronic data acquisition error, and tubing volume error. Following each xenon addition, the heating resistor R in the bottom of the detector operates at 42 W to gasify the liquid, and the whole circulation purification system is allowed to circulate for at least 5 hours to ensure full mixing. It was shown in the literature~\cite{soluble} that xenon is soluble in liquid argon at 87 K up to 16$\%$ by weight without any problem. During the process of circulating purification, if xenon freezes on the cold head, it will block the circulation path and cause changes in flow rate or pressure in detector. This phenomenon has not been observed in the our experiments.

Two radioactive sources, $^{22}$Na and $^{83{\rm m}}$Kr, have been used to calibrate the detector. The $^{22}$Na source is placed right next to the dewar, which is collimated by a 13 cm thick lead collimator with a hole of 10 mm diameter and aimed at the center of the active target. The $^{22}$Na source decays with a 1.275 MeV $\gamma$ and a positron. Annihilation of a positron produces a pair of back-to-back 511 keV $\gamma$-rays, which are detected by the detector and a CsI crystal at the same time. The coincident signals from the detector and CsI are used to form the trigger and an oscilloscope is used to record the signal waveform from the PMT. The $^{83{\rm m}}$Kr source is used to calibrate the detector at xenon concentration of 10010 ppm. The $^{83{\rm m}}$Kr comes from decay of $^{83}$Rb, which is dispersed into a zeolite trap. During the $^{83{\rm m}}$Kr injection process, $^{83{\rm m}}$Kr gas in $^{83}$Rb trap is then entrained in argon flow and introduced to the detector along with the circulating gas. Details of the $^{83{\rm m}}$Kr manufacture are explained in~\cite{Kr83}. 

All the above mentioned processes including filling, krypton gas injection and purification are controlled by a PID device so that the system can run autonomously and stably.

\begin{figure}[htbp]
  \centering
  \includegraphics[width=9cm]{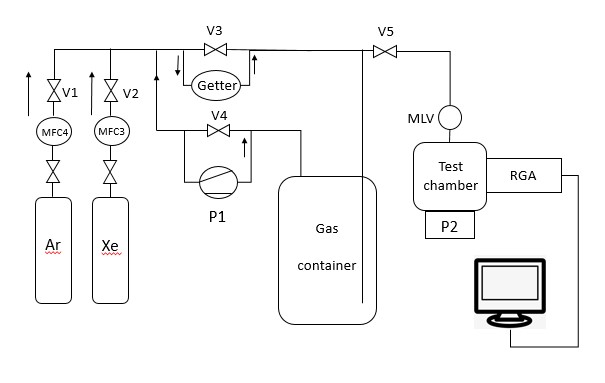}
  \caption{\label{fig:RGA_system}  Schematic view of the RGA calibration system. P1 and P2 are circulation pump and vacuum pump respectively.}
  \end{figure}  
  
\subsection{RGA calibration system}

The concentration of xenon calculated by the MFC is calibrated by the RGA. A schematic view of the calibration setup is shown in Fig.~\ref{fig:RGA_system}. Wherein the principle of a RGA (SRS-RGA200) composed of ionizer, the quadrupole filter and the ion detector is that a small fraction of the residual gas molecules are ionized (positive ions), then the resulting ions are separated according to their mass-to-charge ratios, and the ion currents at each mass are measured. Considering that the RGA is calibrated using N$_{2}$ gas at the manufacturer, argon and xenon have different detection sensitivities from N$_{2}$. The partial pressures recorded in the RGA software do not represent the exact amount of each gas, but the ratio of which differ by a scale factor at different mixing ratios. Therefore, a calibration method based on whether the ion current ratio of xenon to argon obtained by the RGA has a linear relationship with the variation of xenon-doped concentration to simply determine the accuracy of the xenon concentration calculated by the MFC is proposed.

\begin{figure}[htbp]
  \centering
  \includegraphics[width=6.7cm]{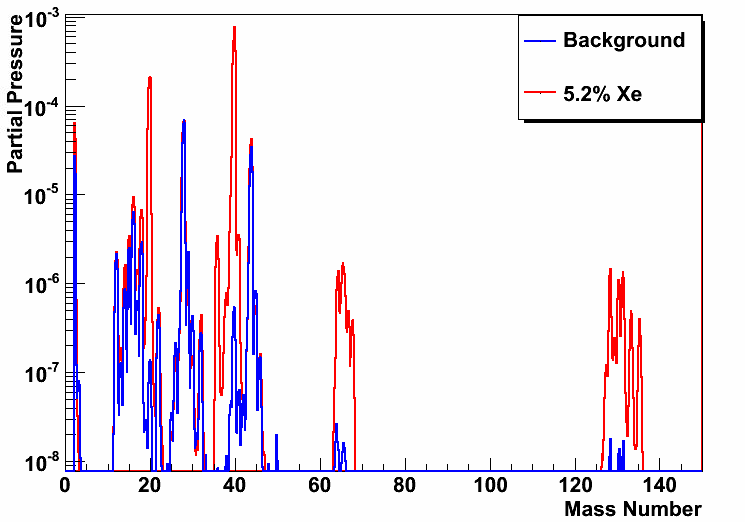}
  \qquad
  \includegraphics[width=7.5cm]{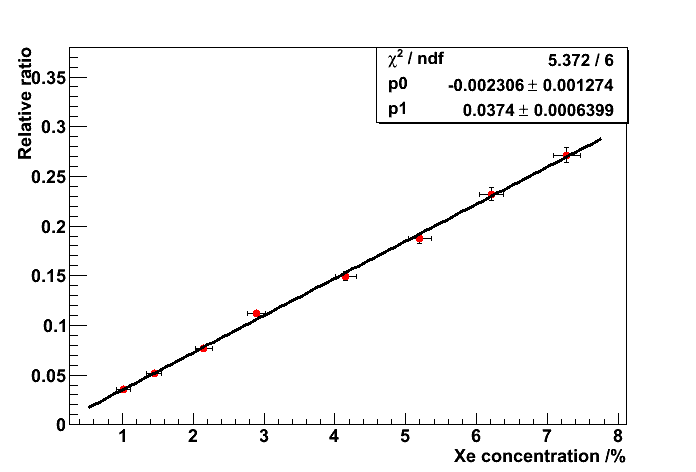}
  \caption{\label{fig:RGA_response}  Left: The mass spectrum of the RGA response for background and 5.2$\%$ xenon-doped argon. The corresponding peak positions with mass numbers equal to 18, 32, 40, 44 and 130 are H$_{2}$O, O$_{2}$, Ar, CO$_{2}$ and Xe, respectively. Mass numbers equal to 20 and 65 are attributed to double ionized argon and double ionized xenon, respectively. Right: The partial pressure ratios of xenon to argon measured by the RGA vs different xenon concentrations calculated by the MFC. The red dots are experimental data and the black line is the liner fitting.}
  \end{figure}
  
We prepared xenon argon mixture samples in the gas container with concentrations varying from 1$\%$ to 8$\%$ at room temperature. Before commencing the measurements the test chamber was baked at 70$\,^{\circ}\mathrm{C}$ and pumped for five days in order to reduce the background caused by the outgassing. The stable residual pressure of the test chamber was about 2*10$^{-10}$ bar, and the mass spectrum of the residual gas was measured by the RGA, which could be used as background spectrum. The mixture sample flows into the test chamber through the variable leakage valve until the pressure in the chamber reaches 2.5*10$^{-7}$ bar $\sim$ 3*10$^{-7}$ bar. The mixture in the test chamber could be used as a sample for RGA measurement. Before each sample measurement, the operation of evacuating the test chamber to about 2*10$^{-10}$ bar and measuring the mass spectrum of RGA response for background could be repeated. Figure~\ref{fig:RGA_response} left shows the mass spectrum comparison of the RGA response for background and 5.2$\%$ xenon concentration. The partial pressure in the mass spectrum displayed by the RGA is calculated in the RGA software by measuring the intensity of the ion current. Figure~\ref{fig:RGA_response} right shows the relationship between the partial pressure ratios of xenon to argon and different xenon concentrations calculated by the MFC. A linear function can be used to describe the relationship perfectly by fitting the experimental data. From this linear relationship, it can be known that the xenon-doped concentration calculated by the MFC is accurate.

The above calibration result also demonstrate the feasibility of using the RGA to directly measure the ratio of xenon to argon after evaporation of liquid mixture to ensure that the xenon does not freeze on the cavity wall or the cold head. We propose a method that can directly measure the xenon concentration in the liquid mixture using the RGA in the our future work, that is, a fine straight tube with a diaphragm valve is used to extract the liquid mixture, one end of which is directly inserted into the liquid mixture, and the other end is connected to a buffer container evacuated. When the diaphragm valve is opened, the liquid mixture will escape from the fine tube under the operating pressure into the buffer container and evaporate. The xenon-doped concentration can be obtained by measuring the ratio of xenon to argon in the buffer container using the RGA.
  
\section{Analysis and results}

\subsection{PMT calibration}

\begin{figure}[htbp]
  \centering
  \includegraphics[width=11cm]{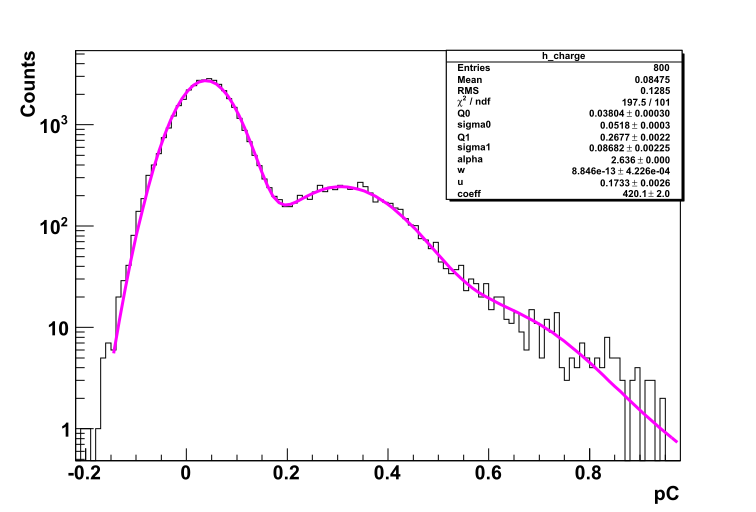}
  \caption{\label{fig:PMT_calibration} The single p.e. spectrum of was obtained by LED calibration. The spectrum was fitted using a PMT response function described in ~\cite{PMTCalib}.}
  \end{figure}

The PMT works at -1400 V, and its gain is calibrated via a LED. Signals of the PMT are recorded by a LeCroy digitizing oscilloscope (HDO6054). The single p.e. spectrum of the PMT is shown in Fig.~\ref{fig:PMT_calibration}, fitted using a PMT response function described in~\cite{PMTCalib}. From the spectrum we can get the mean charge for a single photon-electron which is about 0.27 pC. Using the mean charge, the light yield of the detector can be calculated when the detector is calibrated with a radioactive source.

\subsection{Scintillation signal}

\subsubsection{Waveform model}

When a particle interacts with an argon atom, the excited atoms (excitons, four excition states, $^{1}P_{1}, ^{3}P_{0}$, $^{3}P_{1}, ^{3}P_{2}$) and the ionized atoms (ions) will be produced. Argon atoms would not usually form stable molecules, but in certain states of excitation ($^{3}P_{1}, ^{3}P_{2}$), or when ionized, they can form strong bonds with a ground state argon atom, leading to excited dimers (so called excimers) Ar$_{2}^{*}$ and to ionized dimers Ar$^{+}_{2}$ which undergo the recombination process to form excimers. Most of the scintillation photon of liquid argon come from argon excimers Ar$_{2}^{*}$, which exist as either singlet state or triplet state, then de-excitate with decay time of 7 ns and 1.6 $\mu$s~\cite{1983}, respectively. Since the photons from the excimer decay do not have enough energy to be resonance absorbed by argon atoms, argon is  self-transparent to its own scintillation~\cite{thesis}. The configuration of xenon outer shell electrons is the same as that of argon, therefore it will also form excimers in a singlet and in a triplet state. Liquid xenon scintillates at center wavelength of 176 nm and decay times of different excimer states are 4 ns and 22 ns~\cite{1983}. 

In order to construct the waveform model, the possible physical processes that take place in the scintillator medium and cause the scintillation will be described herein. When doping xenon into argon, it is generally believed that energy can transfer from argon excimers Ar$_{2}^{*}$ including singlet state and triplet state~\cite{2014,2019,2017-1,1993-4} to xenon atoms to form xenon excimers through collisions or dipole-dipole interaction. There are still some doubts in the energy transfer process, such as whether there is an intermediate molecular state (ArXe)$^{*}$ which may play a role in the energy transfer~\cite{1993-1}. Some experimental groups believe that (ArXe)$^{*}$ molecular de-excitation will emit infrared light, which has also been observed in the literature~\cite{waveformlength}. When considering the existence of (ArXe)$^{*}$ molecular, the widely used mechanism of energy transfer~\cite{2014,2019,1993-4} is described as follows:

\begin{equation}
\begin{aligned}
\label{equ:ArXe}
Ar^{*}_{2}(^{1,3}\Sigma^{+}_{\mu}) + Xe + (migration) &\rightarrow (ArXe)^{*} + Ar  \\
 (ArXe)^{*}  + Xe + (migration) &\rightarrow Xe^{*}_{2}((^{1,3}\Sigma^{+}_{\mu}) + Ar
 \end{aligned}
\end{equation}

Meanwhile, mechanism of energy transfer may involve the process where the argon excimers  transfer energy to a xenon atom to excite it, and its excited states($^{1}P_{1}, ^{3}P_{0}$) form lower level excited states ($^{3}P_{1}, ^{3}P_{2}$) after collision, which can form xenon excimers with a ground state atom. The above physical process can be discribed as follows~\cite{1996,thesis}:  

\begin{equation}
\begin{aligned}
\label{equ:Xe}
Ar^{*}_{2}(^{1,3}\Sigma^{+}_{\mu}) + Xe + (migration) &\rightarrow Xe^{*} + 2Ar  \\
 Xe^{*} + Xe + (migration) &\rightarrow Xe^{*}_{2}((^{1,3}\Sigma^{+}_{\mu})
\end{aligned}
\end{equation}

Previous literature studies have observed the enhanced ionizaiton yield for xenon-doped liquid argon~\cite{1976}, which may be attributed to the ionizing excitation transfer process from argon excitons to xenon through dipole-dipole interaction. The possible physical processes that cause scintillator to appear also include that xenon will be directly excited and ionized, although the number of xenon atoms is small relative to argon atoms. The energy transfer as described in equation~\ref{equ:Xe} can as well occur after the argon excimer decay, if a xenon atom absorbs the released scintillation photon.


The light production rate is the sum of the emission rates from the argon and xenon excimer emission rates. It is well known that the waveform model in the xenon-doped liquid argon is much more complicated than that of pure argon, because the above-mentioned physical processes leading to the production of argon and xenon excimer may exist in the medium at the same time. Taking into account all possible physical process would lead to an unwieldy equation and to a large number of free parameters. A very crude analysis might be approximately right for most of the concentrations involved here, by ignoring some physical processes including the ionizing excitation transfer process from argon excitons to xenon and interaction process between xenon atoms and incident particles. 

From the physical processes described by equations~\ref{equ:ArXe}, \ref{equ:Xe}, we know that the processes from the argon excimer to the xenon excimer usually go through two energy transfer steps. To be simple, we use $\lambda_{m}$ to represent the effective rate constants of the first step in the three physical processes including the conversion of the argon excimer into the combined argon-xenon excimer or xenon exciton and use $\lambda_{d}$ to represent the effective rate constants of the second step in the three physical processes including the conversion of the argon-xenon excimer or xenon exciton into xenon excimer. Although the singlet or triplet state of argon excimer may transfer energy through collisions or dipole-dipole interaction, the singlet is more likely to tansfer energy through dipole-dipole interaction due to the short lifetime. It's apparently that the rate constants of energy transfer between singlet and triplet are different. When xenon-doped concentration is small, the argon excimer in the energy transfer is mainly triplet, and when the concentration is increased, the phenomenon of singlet participation will become obvious which has observed in previous literature studies~\cite{2019}. The results in the literature show that the rate constant of the triplet is about twice that of the singlet, but at the same time, the rate constant of the singlet has a large error. 

To reduce the waveform model parameters here, we assume that the transfer rate is the same for both singlet and triplet transfers. The differential equations describing the transfer process between different states can be expressed as in the literature~\cite{2014}. It should be noted that the literature~\cite{2014} only considers the conversion of the argon excimer into the combined argon-xenon excimer, and other energy transfer processes have not been described. Solving these equations, the total light output rate in literature~\cite{2014} is as follows:

\begin{equation}
\label{equ:fit}
r = A_{1}e^{-\frac{t}{T_{f}}} +  A_{2}e^{-\frac{t}{T_{s}}} -  A_{3}e^{-\frac{t}{T_{d}}}
\end{equation}
where $\emph{T$_{f}$}$ and $\emph{T$_{s}$}$ are the decay time of the fast and slow components of scintillation, respectively. $\emph{T$_{d}$}$ is energy transfer time. These three parameters can be described by the following formula:

\begin{equation}
\label{equ:parameters}
T_{f} = \frac{1}{\lambda_{m}+\lambda_{Ar,1}}\quad   T_{s} = \frac{1}{\lambda_{m}+\lambda_{Ar,3}}\quad   T_{d} = \frac{1}{\lambda_{d}}
\end{equation}
where ${\lambda_{Ar,1}}$ and ${\lambda_{Ar,3}}$ represent the decay times for the singlet and triplet in case of pure argon. 

However, the waveform model in many xenon-doped liquid argon experments has not considered the time profile of TPB emission. It is very difficult to measure the time response of TPB at 128 nm photons from liquid argon scintillation or 176 nm photons from liquid xenon scintillation due to the requirement of a fast pulsed light source with corresponding wavelength. Some research groups~\cite{TPB-1} have measured the time response of TPB to 128 nm photons based on the features of the liquid argon scintillation light itself and in particular of the fact that it can be reduced to very fast pulse if the liquid is heavily contaminated by Nitrogen. It is pointed out in the literature~\cite{TPB-1} that the function made of four decaying exponentials can be used to characterize the time evolution of photons after TPB absorbs liquid argon scintillation and re-emitted. The first two exponentials, namely the instantaneous component and the intermediate component have a photon abundance of about 90$\%$. Therefore, simple treatment in this research work, only these two components, that is, the fast and slow components of TPB mentioned below, will be considered. 

According to the datasheet of the PMT, the efficiency of PMT at 176nm is about 5$\%$ lower than the efficiency at 420nm corresponding to the peak of the TPB emission spectrum and the efficiency of PMT at 128nm is almost zero. In addition, the TPB wavelength conversion efficiency obtained in the literatures~\cite{TPB-2, TPB-3, TPB-4} is high, so the contribution to the light yield caused by the 176nm photon directly passing through the TPB and hitting the PMT can be ignored.

Taking this into account, we propose a new waveform model by convolving equation~\ref{equ:fit} with TPB reponse function made of two exponential function. 

\begin{equation}
\label{equ:fit_TPB}
r = (A_{1}e^{-\frac{t}{T_{f}}} +  A_{2}e^{-\frac{t}{T_{s}}} -  A_{3}e^{-\frac{t}{T_{d}}}) \otimes (A_{4}e^{-\frac{t}{T_{f, TPB}}} +  A_{5}e^{-\frac{t}{T_{s, TPB}}})
\end{equation}
where $\emph{T$_{f, TPB}$}$ and $\emph{T$_{s, TPB}$}$ represent decay times for the fast and slow components of TPB time response, respectively. Using equation \ref{equ:fit_TPB} to fit the waveforms of different xenon-doped concentrations can obtain the time characteristics of the energy transfer process and the decay times of the fast and slow components of TPB time response.

\subsubsection{Average waveform shapes}

  \begin{figure}[htbp]
  \centering
  \qquad
  \includegraphics[width=11cm]{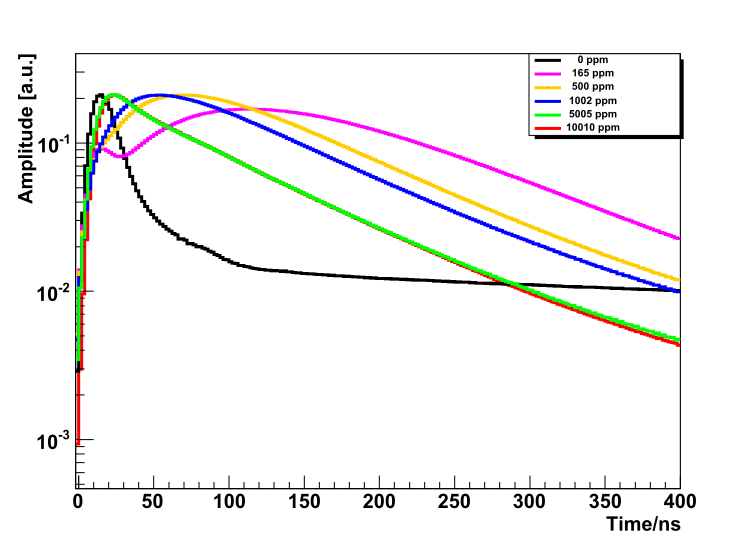}
  \caption{\label{fig:waveform} The comparison of avarage waveform shapes from the PMT at different xenon concentrations.}	
  \end{figure}
  
At each xenon concentration, the signals coming from PMTs in liquid argon detector and in CsI crystal triggered by the back-to-back 511 keV $\gamma$ source pass through a low threshold discriminator module with threshold of 4 mV and 7 mv, separately. A coincident signal created in a logic unit is used as a trigger input for the oscilloscope which record the signal from liquid argon detector with a window of 10 $\mu$s. The waveforms were aligned at the trigger time, and then averaged over at least 10$^{5}$ events. The time interval between each concentration measurement is at least one day. Average waveform shapes at different concentrations are shown in Fig.~\ref{fig:waveform} with 400 ns time window.

 \begin{figure}[htbp]
  \centering
  \subfigure[0 ppm]{
  \includegraphics[width=7cm]{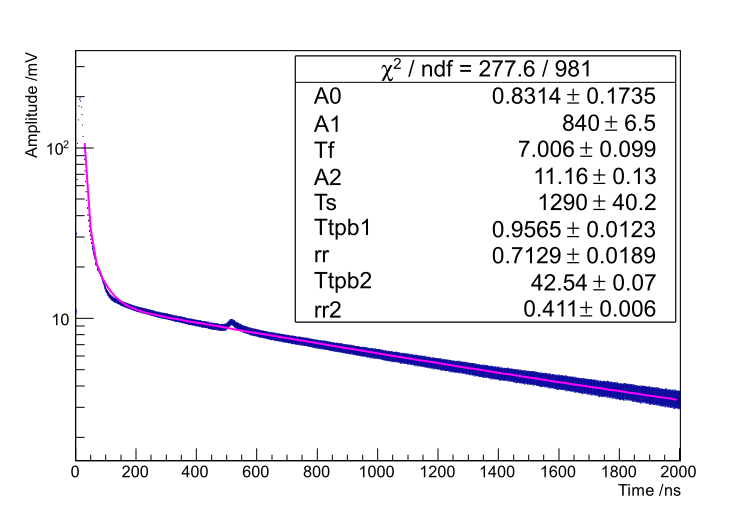}
  }
  \quad
  \subfigure[165 ppm]{
  \includegraphics[width=7cm]{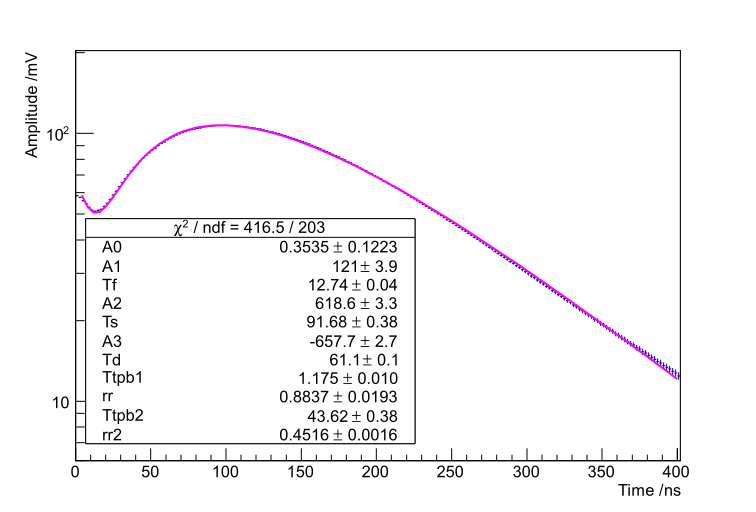}
  }
  \quad
  \subfigure[500 ppm]{
  \includegraphics[width=7cm]{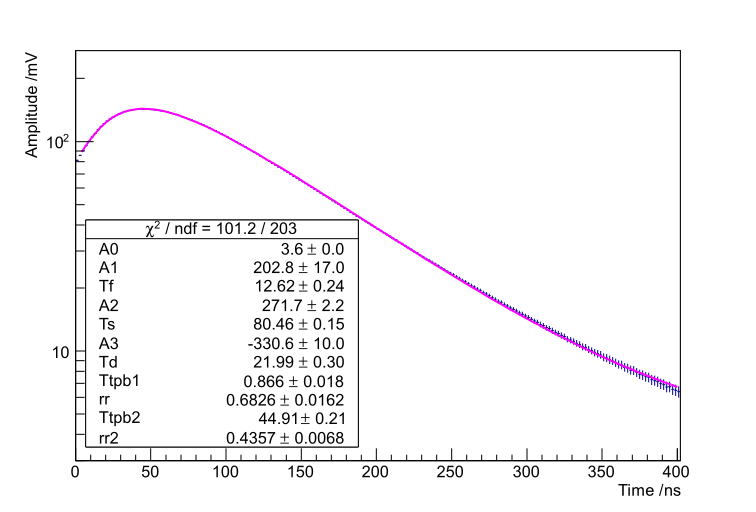}
  }
  \quad
  \subfigure[1002 ppm]{
  \includegraphics[width=7cm]{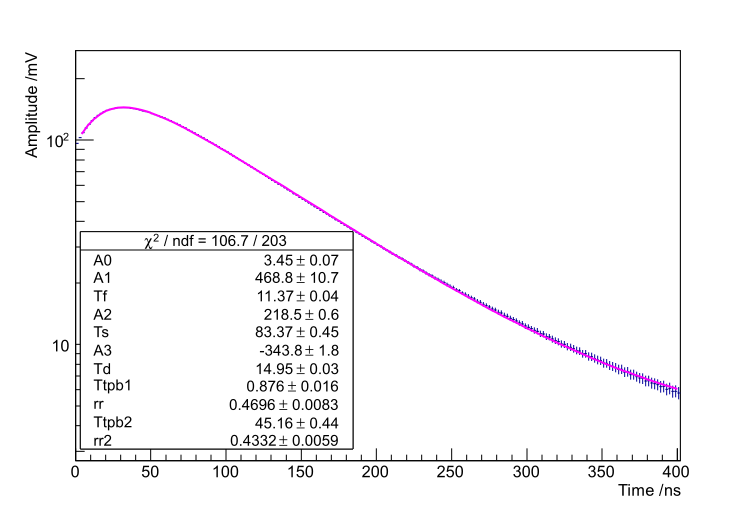}
  }
  \quad
  \subfigure[5005 ppm]{
  \includegraphics[width=7cm]{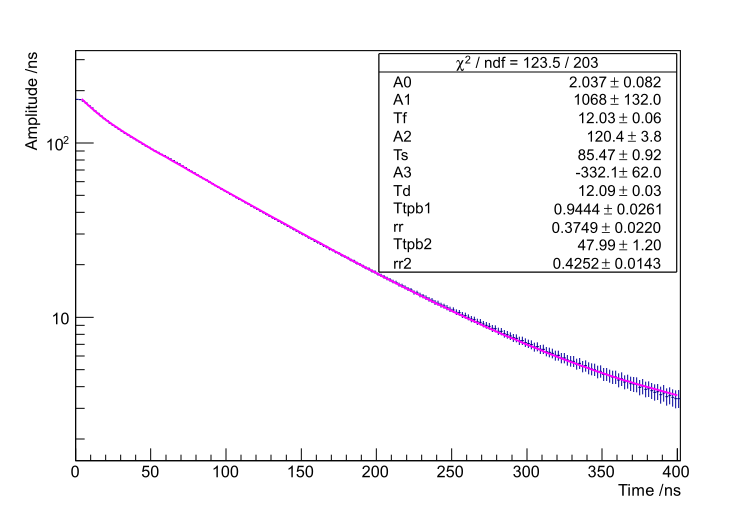}
  }
  \quad
  \subfigure[10010 ppm]{
  \includegraphics[width=7cm]{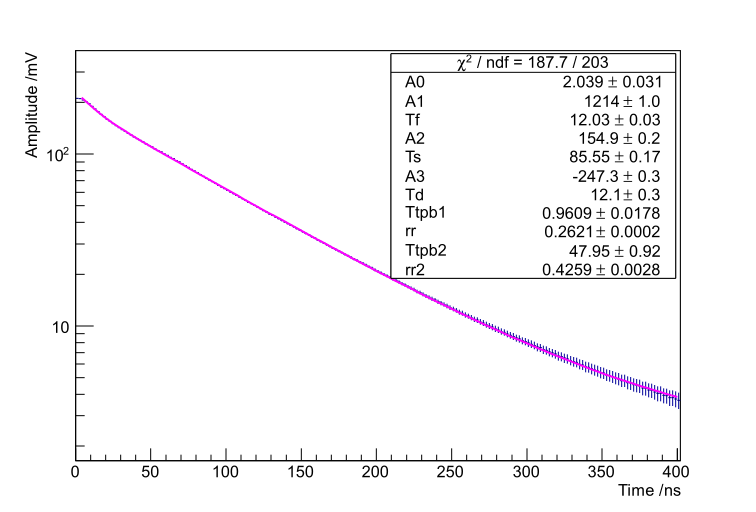}
  }
  \caption{\label{fig:fit_waveform} Fitting average waveform at different xenon-doped concentrations. }	
  \end{figure}
  
In pure liquid argon, the waveform is marked as 0 ppm in the Fig.~\ref{fig:waveform} which obviously has a fast and slow exponential decay component. When the concentration of xenon-doped gradually increases, the decay time of the slow component will gradually decrease. The signal results in a widened waveform, and the waveform will gradually form two humps due to the presence of the fast component of argon, as shown by the waveform of 165 ppm in Fig.~\ref{fig:waveform}. The waveform distortion is consistent with results in the literatures~\cite{2014,2017,2019}. The first hump is caused by the scintillation of argon and the second hump is the consequence of argon excimers transferring to xenon excimers. In addition, the results in the literature~\cite{2014} show that even if the xenon-doped concentration is about 9 ppm, there is already a sign of two humps in the waveform. For comparison, it was found by using a monochromator that even if the xenon-doped concentration is about 1 ppm in the literature~\cite{waveformlength}, the scintillation of xenon has increased a lot compared with that of pure argon.

As doping continues, the second hump moves backward while the first hump decreases in intensity, which is due to the increase in the ratio of energy transfer of the singlet of argon when the xenon-doped concentration becomes higher, this phenomenon has been confirmed in the literature~\cite{2019}. When the concentration increases to several hundred ppm (may be less than 500 ppm), the two humps will merge together and the merged hump will continue to move backward. It can be observed in the Fig.~\ref{fig:waveform} that when the concentration is less than or equal to 5005 ppm, the energy transfer is almost complete.

The scintillation signal parameters were obtained by using the waveform model described by equation~\ref{equ:fit_TPB} to fit waveform shapes at each concentration. The waveform fitting results are shown in Fig.~\ref{fig:fit_waveform}. For comparison, the convolution function of the double exponential function, assumed to be time profile of liquid argon scintillation, with the response function of TPB is used to fit the waveform of pure argon. Because there is a non-zero baseline in the waveforms, a positive constant baseline term is also added to as an additional parameter in the fit. It can be seen in Fig.~\ref{fig:fit_waveform} (a) that a hump appears on the waveform at a time of around 500 ns, which may be an afterpulse caused by the residual gas in the PMT. For pure argon case, the slow component time constant is much greater than 500 ns, so the impact of the hump structure on the time constant is not considered. For xenon-doped liquid argon cases, whether the fitting range includes this feature will affect the parameter values obtained. The fitting ranges for pure argon and xenon-doped liquid argon are 2000 ns and 400 ns, respectively. The parameters in fitting function are assigned initial values, and then free fitting is performed without parameters restriction. Changing the fit region slightly produced changes in the parameters of less than 5$\%$. The fit matched the data quite well. The decay times ($\emph{T$_{f}$}$ and $\emph{T$_{s}$}$) for the fast and slow components of the scintillation, the time ($\emph{T$_{d}$}$) to describe the energy transfer processes and the decay times ($\emph{T$_{f, TPB}$}$ and $\emph{T$_{s, TPB}$}$) for the fast and slow components of TPB time response are shown in table~\ref{tab:decay_time}. These results only quote the statistical errors from the fit. Since the fitting function does not include the response of the detector itself, the effects of photon propagation process, electronics, PMT, etc. on time are not considered. The errors in the table do not include errors due to the response of the detector itself. On all waveforms, insufficient time resolution caused by the reponse of the detector and presence of electronic noise result in quite large uncertainties in these parameters.

\begin{table}[]
\centering
\resizebox{\textwidth}{!}{%
\begin{tabular}{l|c|c|c|c|c|c|c}
\hline
 & Yield (p.e./keV)  & $\emph{T$_{f}$}$ (ns) & $\emph{T$_{s}$}$ (ns) & $\emph{T$_{d}$}$ (ns) & $\emph{T$_{f, TPB}$}$ (ns) & $\emph{T$_{s, TPB}$}$ (ns) & Slow/Fast\\ 
\hline
Pure argon & 8.4 $\pm$ 0.4 & 7.01 $\pm$ 0.10 & 1290.0 $\pm$ 40.2 &  --- & 0.96 $\pm$ 0.01 & 42.54 $\pm$ 0.07 & 0.41 $\pm$ 0.01 \\
165ppm xenon & 9.9 $\pm$ 0.5 & 12.74 $\pm$ 0.04 & 91.68 $\pm$ 0.38 & 61.1 $\pm$ 0.1 & 1.18 $\pm$ 0.01 & 43.62 $\pm$ 0.38 & 0.45 $\pm$ 0.01  \\
500ppm xenon & 9.8 $\pm$ 0.5 & 12.62 $\pm$ 0.24 & 80.46 $\pm$ 0.15 & 21.99 $\pm$ 0.30 & 0.87 $\pm$ 0.02 & 44.91 $\pm$ 0.21 & 0.44 $\pm$ 0.01\\
1002ppm xenon & 9.7 $\pm$ 0.5 & 11.37 $\pm$ 0.04 & 83.37 $\pm$ 0.45 & 14.95 $\pm$ 0.03 & 0.88 $\pm$ 0.02 & 45.16 $\pm$ 0.44 & 0.43 $\pm$ 0.01 \\
5005ppm xenon  & 9.7 $\pm$ 0.6 &  12.03 $\pm$ 0.06 & 85.47 $\pm$ 0.92 & 12.09 $\pm$ 0.03 & 0.94 $\pm$ 0.03 & 47.99 $\pm$ 1.20 & 0.43 $\pm$ 0.01 \\
10010ppm xenon  & 9.8 $\pm$ 0.6 & 12.03 $\pm$ 0.03 & 85.55 $\pm$ 0.17 & 12.10 $\pm$ 0.30 & 0.96 $\pm$ 0.02 & 47.95 $\pm$ 0.92 & 0.43 $\pm$ 0.01 \\
\hline
\end{tabular}%
}
\caption{Time parameters of scintillation in liquid argon, xenon-doped liquid argon at different concentrations and TPB.}
\label{tab:decay_time}
\end{table}

For different concentration results, $\emph{T$_{f}$}$ is almost the same within the error range, and it is consistent with the reported values of about 10 ns in the literatures~\cite{2014,2019}. $\emph{T$_{s}$}$ decreases with increasing concentration and eventually reaches a about 85 ns platform, which is lower than the reported values of about 100 ns in the literatures~\cite{2014,2019}. The reason for causing this difference is that considering the decay time of TPB would reduce $\emph{T$_{s}$}$. $\emph{T$_{d}$}$ has the same trend like $\emph{T$_{s}$}$, but the decreasing trend of $\emph{T$_{d}$}$ is more obvious as the concentration increases. The platform value of $\emph{T$_{d}$}$ is almost consistent with the reported value at the concentration of 1000 ppm in the literature~\cite{2014}.

One can see from the table \ref{tab:decay_time} that $\emph{T$_{f, TPB}$}$ and $\emph{T$_{s, TPB}$}$ are almost same at different concentration, even for pure argon. It should be noted that because of the influence of TPB and purity of the liquid argon, the value $\emph{T$_{s}$}$ of pure argon is lower than 1.6 $\mu$s. The above phenomenon is also mentioned in the literature~\cite{waveformlength}, in  which measured value is about 1.3 $\mu$s, which is consistent with the our results. In addition to the decay times for the fast and slow components of TPB, the relative abundance ratios of the slow to fast components are also listed in the table at different concentrations, and their values agree with the results in the literature~\cite{waveformlength} which uses liquid argon scintillation quenched by nitrogen contaminations, $\beta$ and $\alpha$ particles to excite TPB. These results demonstrate the feasibility of extracting time parameters of TPB from waveform in pure argon or xenon-doped liquid argon. 

\begin{figure}[htbp]
  \centering
  \includegraphics[width=8cm]{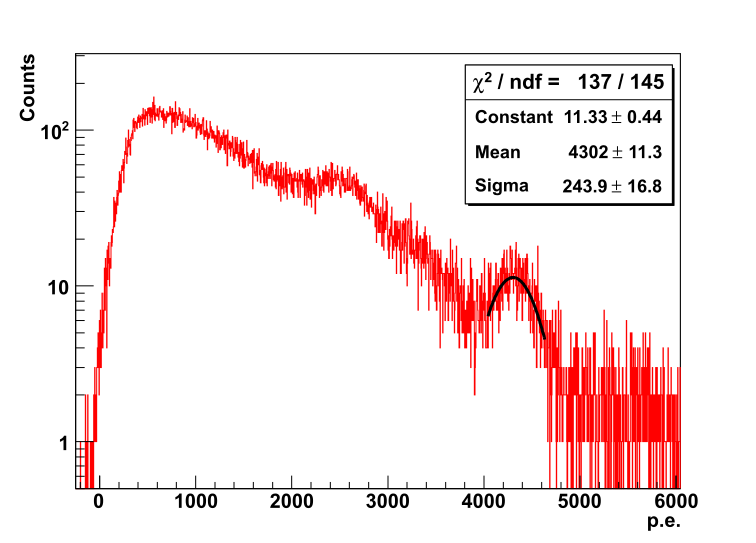}
  \caption{\label{fig:Naspectrum} The energy spectrum of $^{22}$Na and the black line is the fitting result of the full energy peak of the 511 keV $\gamma$.}
  \end{figure}

\subsection{Light yield }

The light yield is calculated from the number of photoelectrons generated by the PMT corresponding to the full energy peak of the 511 keV $\gamma$ energy spectrum. For pure argon case, the energy spectrum for $^{22}$Na is shown in Fig.~\ref{fig:Naspectrum}. The spectrum is fitted with a Gaussian function. The light yields at each concentration are shown in table \ref{tab:decay_time}. One could observe an increase of light yield from (8.4 $\pm$ 0.4) p.e./keV to (9.9 $\pm$ 0.5) p.e./keV at concentrations of 0 ppm and 165 ppm. A conclusive explanation for the light yield increase has not been given yet either.It may be that the xenon-doped does seem to lower the W-value which is the average energy needed to produce one electron-ion pair and a higher quantum efficiency of the TPB wavelength shifter at the emission wavelength of xenon excimers~\cite{thesis}. The light yield may have reached saturation at 165 ppm or less than it in our work, however, a possible plateau is observed in the signal yield at about 100 ppm in the results in~\cite{2014} which explains that the maximum fraction of excimer states are being transferred to xenon at this concentration. The conclusion of another experimental group~\cite{2019} shows that the light yield saturates at the concentration (may be larger than 590 ppm) of which reaches the point of the total re-emission of the fast component by xenon. The inconsistency of the results between the different experimental groups has not been understood yet, which may be related to the detector configuration or the physical processes considered.

  \begin{figure}[htbp]
  \centering
  \includegraphics[width=7cm]{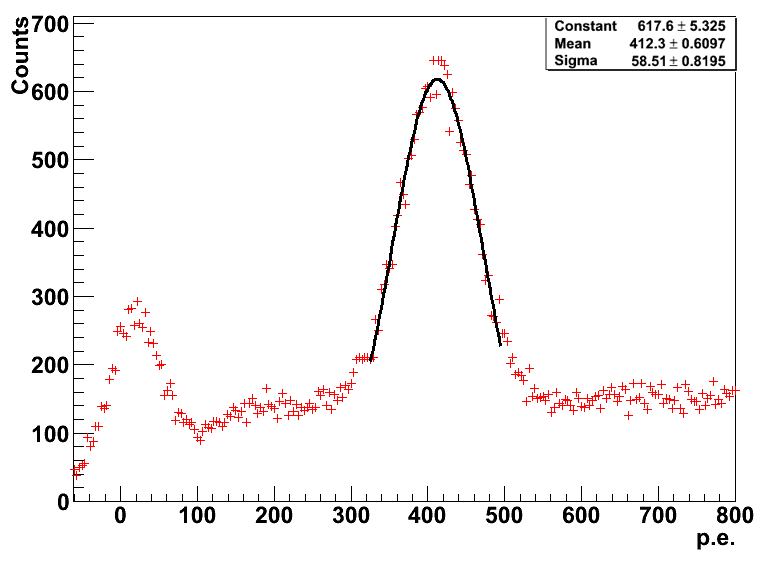}
  \qquad
  \includegraphics[width=7cm]{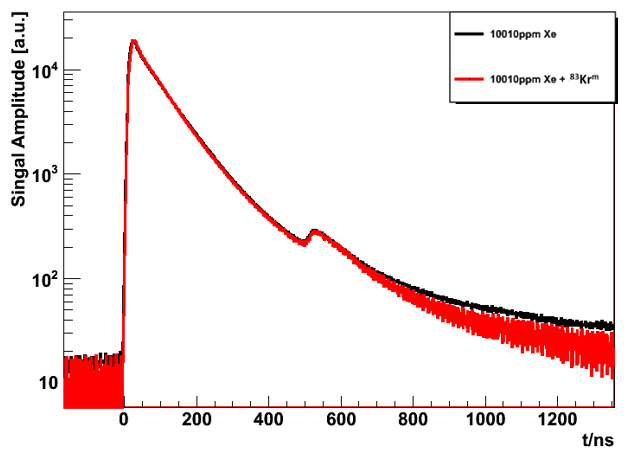}
  \caption{\label{fig:spectrum_Kr} The spectrum of $^{83{\rm m}}$Kr and waveform comparison in the presence or absence of $^{83{\rm m}}$Kr.}
  \end{figure}

\subsection{$^{83{\rm m}}$Kr response}

The $^{83{\rm m}}$Kr could be used for calibration to obtain the scintillation efficiency difference over the range of 41.5 keV to 511 keV. At the same time, considering the fast and slow components of liquid krypton scintillation photons have decay times of about 2 ns and 90 ns respectively~\cite{krypton}, the effect of the introduction of trace amounts of $^{83{\rm m}}$Kr~\cite{1972} on the waveform can be verified.

The energy spectrum of 41.5keV $\gamma$ emitted from $^{83{\rm m}}$Kr at xenon concentration of 10010 ppm is shown in Fig.~\ref{fig:spectrum_Kr} left, where the energy spectrum does not subtract the background spectrum. A light yield of (9.9 $\pm$ 1.4) p.e./keV is achieved, with an energy resolution of (14.5 $\pm$ 0.2)$\%$ by fitting the full energy peak using a gaussian function. As a comparison, light yield for 511 keV gamma is (9.8 $\pm$ 0.6) p.e./keV, with an resolution of (5.3 $\pm$ 0.2)$\%$. There is no systematic difference between the signal yields measured at the different energies, from which we infer that the scintillation efficiency is almost same over the range of 41.5 keV to 511 keV. The Figure~\ref{fig:spectrum_Kr} right shows waveform comparison in the presence or absence of $^{83{\rm m}}$Kr. The waveforms in both cases almost overlap at the time less than 800 ns. However, a small difference can still be observed when the time is greater than 800 ns. This difference has tiny effect on the calibration results and can be ignored. 

\section{Conclusion and discussion}

In this work the trend of average waveform shapes and light yield with different xenon concentrations were studied by a collimated $^{22}$Na source. To ensure that the concentration calculated by the MFC is accurate, the RGA was used to calibrate it at different concentrations.

Many physical processes are considered and reviewed, and a new waveform model based on the described physical processes is established to fit the PMT waveform at different xenon concentrations. As xenon doping into liquid argon, two humps will appear on the waveform, and they will merge together at high concentration, wherein the second hump is due to energy transfer from argon eximers to xenon eximers. The decay times for the fast and slow components of the scintillation, the time of energy transfer process, and the decay times for the fast and slow components of TPB can be extracted by fitting the waveform. Most of the fitting results are consistent with those in the literature, except that the decay time for slow component becomes smaller due to considering the decay time of TPB. Xenon-doped liquid argon will also cause an increase in light yield, which may be due to lower the W-value or shifting of the scintillation wavelength~\cite{thesis}. 

Moreover, $^{83{\rm m}}$Kr is successfully introduced into the xenon-doped liquid argon detector for the first time. From results of $^{83{\rm m}}$Kr calibration, it could be known that the scintillation efficiency is almost same over the range of 41.5 keV to 511 keV. The impact of the introduction of trace amounts of $^{83{\rm m}}$Kr source on the waveform is mainly reflected in the part of time greater than 800 ns, but the impact is very small.

\acknowledgments

This work was supported by the science and technology innovation project of Institute of High Energy Physics Chinese Academy of Sciences (2017IHEPZZBS116). The author acknowledges Dr. Y. Wang for the discussions on preparing xenon argon mixture.


\end{document}